\documentstyle[12pt,epsf]{article}
\let\section=\subsection     \let\subsection=\subsubsection                
\def\Journal#1#2#3#4{{#1} {\bf #2}, #3 (#4)}

\def\PRA{{\em Phys. Rev.} A}

\def\RMP{{\em Rev. Mod. Phys.}}
\def\PREPC{{\em Phys. Rep.} C}

\newcommand{\be}{\begin{equation}}
\newcommand{\ee}{\end{equation}}
\newcommand{\bea}{\begin{eqnarray}}
\newcommand{\eea}{\end{eqnarray}}

\newcommand{\hf} {{1\over2}}

\def\ra{\rangle}
\def\la{\langle}
\def\hf{{1\over2}}
\def\dk{\Delta k}
\def\eq#1{(\ref{#1})}

\def\mathrm{}

\begin{document}
\begin{center}
   {\large \bf MULTIFRAGMENTATION AND THE RENORMALIZATION GROUP\footnote{Presented at the
XXVII International Workshop on Gross Properties of Nuclei and Nuclear
Excitations, Hirschegg (Austria) January, 1999}}\\[2mm]
J. POLONYI \\[5mm]
   {\small \it  Theoretical Physics Laboratory \\
   Louis Pasteur University, 67084 Strasbourg, Cedex, France \\
   and\\
   Department of Atomic Physics, L. E\"otv\"os University \\
   P\'azm\'any P. S\'et\'any 1/A 1117 Budapest, Hungary \\[8mm] }
\end{center}

\begin{abstract}\noindent
The possible usefulness of the renormalization group (RG) method 
in Nuclear Physics is pointed out in this talk in the context of 
the nuclear multifragmentation. 
The presentation is rather superficial and sketchy, to indicate 
the main lines only. But I believe that the idea is warrant 
of further, more careful studies. 
\end{abstract}

\section{The Renormalization Group}
Motto: "{\em We should rather eliminate a degree of freedom
if it is not playing an important role in a given observable.
The complications left behind, such as the effective interactions, 
form factors, etc. are easier to cope with than the complications related to
a truly dynamical variable.}" A successful implementation of this idea 
can be found in quantum chemistry where the d,f and g electrons are
taken into account by means of the form factors rather than keeping them
as true dynamical degrees of freedom. In an analogous manner, the
application of the RG method for the multifragmentation processes
is based on the effective treatment of the nuclear bound state
structure.

The RG method has two independent bonuses. It (i)
helps to identify those few microscopic coupling constants which characterize
a given dynamics \cite{wilsrg}, and (ii) provides a non-perturbative tool for 
solving quantum field theories.

\underbar{Elementary interactions:}
Suppose that the elementary ("microscopic") description of the
dynamics is given by the hamiltonian $H_k[\psi_N,\Phi_M,\chi_{(A,Z)}]$,
where the fields $\psi_N$ and $\Phi_M$ stand for few low lying nucleons and 
mesons, respectively. The nuclei belong to the field $\chi_{(A,Z)}$.
This hamiltonian is supposed to describe the dynamics up to 
$k=\Lambda\approx$few GeV/c single
particle momentum and it contains interaction terms like
$\bar\psi_N\psi_N\Phi_M$, $\Phi_M^4$,
$\chi_{(A,Z)}^\dagger\psi_p\cdots\psi_p\psi_n\cdots\psi_n$
involving $Z$ proton and $A-Z$ neutron fields and
$\chi^\dagger_{(A_1+A_2,Z_1+Z_2)}\chi_{(A_1,Z_1)}\chi_{(A_2,Z_2)}$.
The local field are convoluted with form factors to form the vertices.

\underbar{RG flow:}
Such a model is not particularly interesting because the large number of
the parameters spoils practicality and the predictive power. The
remedy of the problem is the blocking, or the RG step. It consists of the 
lowering of the cutoff, $k\to k'=k-\dk$ by eliminating particles within the 
momentum shell $k-\dk\le p\le k$ and by constructing the effective 
dynamics governed by $H_{k-\dk}$. The coupling constants appearing in 
the new hamiltonian take new values, $g_n\to g'_n={\cal B}_n(g;k,k')$,
their modification represent the dynamics of the modes eliminated. 
The result is the cutoff dependence of the coupling constants 
$g_n(k)$, the RG flow.

What is the use of the RG flow? Suppose that we
need the expectation value $\la A(p_{obs})\ra$ of an observable
at the momentum scale $p_{obs}$. The contributions of particles
with momentum $p>p_{obs}$ are suppressed in the expectation value.
The basic principle of the RG method is to use the effective dynamics
with cutoff slightly above $p_{obs}$ to obtain the expectation value. 
As a result, the coupling constants $g_n(p)$ express the effective 
strength of the interactions at a given scale.
In order to see the true scale dependence the dimension of the
coupling constant is expressed by the cutoff, $k$, and one works
with dimensionless coupling constants.

\underbar{Universality:}
The blocking step allows us to eliminate a large number of coupling
constants from the hamiltonian. This is achieved by linearizing the blocking
transformation ${\cal B}(g;k,k')$ around a fixed point,
$g^*={\cal B}(g^*;k,k')$. Let us denote the eigenvalues of the
linearized blocking, $\partial_{g_n}{\cal B}_m(g^*;k,k')$ 
by $\lambda_n(k/k')$. Two successive
blocking steps give the relation
$\lambda_n(x)\lambda_n(y)=\lambda_n(xy)$. The only continuous solution of this equation is
$\lambda_n(x)=x^{\nu_n}$.
The parameter $\nu_n$ introduced in this manner is called the critical
exponent. The coupling constants with $\nu<0$ or $\nu>0$
are called irrelevant, or relevant, respectively. This classification
corresponds to the scaling regime, the range of the scale $k$ where the
linearization is reliable. The use of this
property obtained in the linearized level is that it asserts that
the RG flow becomes independent of the initial value of the 
irrelevant coupling constants after leaving the vicinity of a
fixed point\footnote{It is important to bear in mind
that the irrelevance is {\em not} the claim that the coupling constant
is weak. It concerns instead the difference between the running coupling constants
and the fixed point.}. 

There are two other important developments to mention:
\begin{itemize}
\item The kinetic energy is the dominant piece of the hamiltonian at
high energy and the models consisting of free, massless particles
represent perturbative fixed points. 
\item The relevant (irrelevant) coupling constants are the renormalizable
(non-renormalizable) ones. 
\end{itemize}
The result is the universality,
the claim that the initial value of the non-renormalizable
coupling constants at the shortest length scale
can safely be set zero and it is enough
to adjust the renormalizable coupling constants only. 

The weak point of applying the RG method in Nuclear Physics
lies here. One may say that one of the fundamental difficulty 
of Nuclear Physics is the closeness of its characteristic scales
\footnote{This results from the strongness of the coupling constants,
c.f. the separation of the scales by $\alpha$ in atomic physics.}: 
one can not talk about hadronic degrees of freedom
below 1fm and the strong interaction is already negligible beyond few
dozens of fm. The UV and IR "neighbors" of Nuclear Physics,
QCD and QED are better placed because they have longer
scaling regimes which allows the sufficiently strong suppression
of the non-universal interactions and thereby eliminate the
need of the non-renormalizable coupling constants. It remains to be
see if the critical exponents of the non-renormalizable
coupling constants are sufficiently negative in Nuclear Physics
to eliminate enough coupling constants at $\Lambda$.

\underbar{Wegner-Houghton equations:}
In order to follow the RG flow of a large number of coupling constants
one needs a specially powerful version of the RG method. This is based
on infinitesimal RG steps where $\dk/k\ll1$. $\dk/k$ acts
as a new small parameter and the RG equation obtained in the 
one-loop level becomes exact as $\dk/k\to0$ \cite{wh}.

\underbar{Time evolution:} A characteristic feature of the 
multifragmentation processes is that the conservation laws 
play an important role in bringing the system out of equilibrium. 
One encounters similar situation in the experimental studies 
of glassy materials \cite{spgl}. The potential
energy of glasses is an extremely complicated function of the 
microscopic parameters and possesses a large number of minima. In the
quenching experiment one starts at a temperature where the
kinetic energy is well above the peaks of the potential energy.
After thermalization the temperature is suddenly lowered
what makes the system trapped at the closest potential minimum.
The time evolution and the relaxation follow different laws
before and after quenching. This is reminiscent the break-up
step in the multifragmentation because the ergodicity is abruptly 
broken in both cases. The method of the
dynamical renormalization group (DRG) was developed to
deal with such problems and to classify the possible time evolutions 
\cite{drg}. This method would, in addition, provide a systematic
description of the statistical aspects of the multifragmentation.

\underbar{Soft modes:} The detectors pick up long time and long
distance observables. This makes one believe that the soft modes,
the light mesons are particularly important for the 
multifragmentation processes. One might envision the 
nuclei and the nucleons emerging from the multifragmentation
process as being surrounded by the classical meson fields
which are responsible for their interactions. The resulting
picture is similar to the cloudy bag model scenario.

The RG method tailored for the multifragmentation problem
may consists of a TDHF scheme where the actual cutoff 
is time dependent and follows the average kinetic energy 
of the particles.

\section{Mixed Phase}
Another application of the RG strategy which might be interesting
in the study of the nuclear multifragmentation is the description
of the spinodal phase separation. Let us consider the partition
function 
\be
Z_\Lambda(\Phi)=\int D[\phi]e^{-{1\over\hbar}S_\Lambda[\phi]}
\delta\left(L^{-d}\int d^dx\phi(x)-\Phi\right)
\ee
where the Landau-Ginsburg free-energy, $S[\phi]$, corresponds
to the phase with spontaneously broken symmetry,
\be
S_k[\phi]=\int d^dx\left[\hf(\partial_\mu\phi(x))^2+V_k(\phi(x))\right].
\ee
The dimensionless parameter $\hbar$ is introduced to organize the
loop-expansion and $L^d$ is the volume. The mean-field solution predicts 
metastability for $\Phi_{sp}(0)\le|\Phi|\le\Phi_0(0)$, where 
\be
k^2+\partial^2_\Phi V_k(\Phi_{sp}(k))=0,~~~
k^2\Phi_0(k)+\partial_\Phi V_k(\Phi_0(k))=0.
\ee
Consider the 
fluctuations with momentum $p<k_{cr}$ and $|\Phi|\le\Phi_{sp}(p)$. 
Here $k_{cr}$ denotes the onset of the instability, i.e. $\Phi_0(k)>0$ 
for $k<k_{cr}$. The amplitude of these modes increases exponentially 
in time because the inverse propagator 
$G^{-1}(p^2)=p^2+\partial^2_\Phi V_p(\Phi)$ is negative. 

The RG proceeds by the successive elimination of the modes,
and the construction of the effective action
\be
e^{-{1\over\hbar}S_{k-\dk}[\phi_{IR}]}=\int D[\phi_{UV}]
e^{-{1\over\hbar}S_k[\phi_{IR}+\phi_{UV}]},
\label{blck}
\ee
where the Fourier transforms $\tilde\phi_{IR}(p)$ and $\tilde\phi_{UV}(p)$ are
non-vanishing for $p<k-\dk$ and $k-\dk<p<k$, respectively.
The action $S_k[\phi]$ has a local maximum at $\tilde\phi_{UV}(k)=0$ in 
the spinodal unstable region. Since $S_k[\phi]$ is bounded from below 
the spinodal instability is characterized by the appearance of a 
nontrivial saddle point, $\tilde\phi^{sp}_k(x)$, in the blocking \eq{blck}.
The saddle point is the solution of the projection of the 
equation of motion of $S_k$ into the momentum space shell $k-\dk<p<k$.
The blocking relation is
\be
e^{-{1\over\hbar}S_{k-\dk}[\phi_{IR}]}=\sum\limits_\alpha
e^{-{1\over\hbar}S_k[\phi_{IR}+\phi_{k,\alpha}^{sp}]}
\int{dX_{k,\alpha}\mu(X_{k,\alpha})\over\sqrt{\det'
{\partial^2S_k[\phi_{IR}+\phi_{k,\alpha}^{sp}(X)]
\over\partial\phi_{UV}\partial\phi_{UV}}}}
\left(1+O(\hbar)\right),\label{trerge}
\ee
where the summation is over the different saddle points.
$\det'$ denotes the determinant in the subspace of the modes
with momentum $k-\dk<p<k$ which is orthogonal to the
saddle point and $\mu(X_{k,\alpha})$ stands for the integral measure of the 
zero modes $X_{k,\alpha}$.
We find a tree-level renormalization because the saddle point depends on the 
infrared background field, $\phi_k^{sp}=\phi_k^{sp}[\phi_{IR}]$. In the case
of a single saddle point \eq{trerge} reads as
\be
S_{k-\dk}[\phi_{IR}]=S_k[\phi_{IR}+\phi_k^{sp}[\phi_{IR}]]+O(\hbar)
\ee

The numerical computation of the effective action was reported
in ref. \cite{treee} by taking into account the plane wave saddle points.
The conclusion can be generalized by assuming the continuity of the 
coupling constants in the cutoff in the unstable region with the following
result \cite{treek}:
\begin{itemize}
\item The mixed phase extends over the metastable region of the
mean-field solution, i.e. the saddle point is nontrivial not 
only for $|\Phi|\le\Phi_{sp}(k)$ but whenever $|\Phi|\le\Phi_0(k)$.

\item The amplitude of the saddle point $\phi^{sp}_k(x)$ is such that
the background field plus the saddle point sweeps through the whole unstable
region, $max_x|\Phi+\phi^{sp}_k(x)|=\Phi_0(k)$.

\item The effective action is degenerate in the mixed phase, i.e.
it is left unchanged by the fluctuations 
$max_x|\Phi+\phi_{UV}(x)|\le\Phi_0(k)$. The gradient expansion ansatz
\be
S_k=\int d^dx[\hf Z_k(\phi(x))(\partial_\mu\phi(x))^2+V_k(\phi(x))]
\ee
yields a simple quadratic effective action 
\be
S_k^{mixed}=\hf Z_k^{stable}(\Phi_0(k))
\int d^dx\left[(\partial_\mu\phi(x))^2-k^2\phi^2(x)\right]
+O\left({\dk\over k}\right)\label{mixpo}
\ee
in the mixed phase where $Z^{stable}$ is read off from the solution of 
the RG equation in the stable region. Notice that this result is exact,
the only correction being $O(\dk/k)$ which can be arbitrarily small.
\end{itemize}
Notice that the last point implies the Maxwell construction. In fact,
$V_{k=0}(\Phi)$ is the free energy density which is flat for 
$-\Phi_0(0)\le\Phi\le\Phi_0(0)$ according to \eq{mixpo}. 
The tree-level renormalization of the potential
$V(\phi)=-0.05\phi^2+0.2\phi^4/4!$ is depicted in Fig.~1.

It is worthwhile mentioning that the correlation functions can easily 
be obtained in the mean-field approximation. The real time evolution
can, as well, be incorporated since it is actually a saddle point.

\begin{center}
\begin{minipage}{13cm}
	\epsfxsize=9cm
	\epsfysize=6cm
	\centerline{\epsfbox{c1}}
\baselineskip=12pt
{\begin{small}
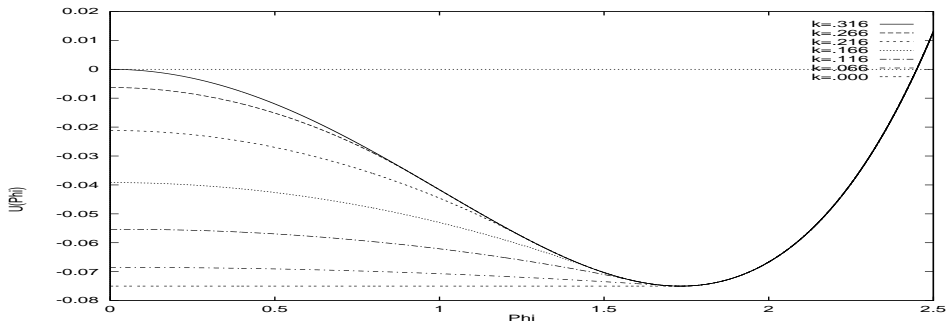
Fig.~1. $V_k(\Phi)$ for different values of $k$ showing the evolution 
towards the Maxwell construction at $k=0$.\end{small}}
\end{minipage}
\end{center}

\end{document}